\documentclass[]{article}
\usepackage{graphicx}

\begin{document}

\centerline{\Large \bf Can a Falling Ball Lose Speed?}
\bigskip
\centerline{P.M.C. de Oliveira$^1$, S. Moss de Oliveira$^1$, F.A.C. 
Pereira$^2$ and J.C. Sartorelli$^2$ }

\vskip10pt\noindent
1) Instituto de F\'{\i}sica, Universidade Federal Fluminense\\
Av. Litor\^{a}nea s/n, Boa Viagem, Niter\'{o}i 24210-340, RJ, Brazil\\
and National Institute of Science and Technology for Complex Systems

\bigskip\noindent
2) Instituto de F\'{\i}sica, Universidade de S\~ao Paulo\\
Rua do Mat\~ao, Cidade Universit\'aria, S\~ao Paulo 05508-090, SP, 
Brazil

\bigskip\noindent
e-mail addresses:\\
pmco@if.uff.br; suzana@if.uff.br; faugusto@if.usp.br; sartorelli@if.usp.br

\bigskip

\begin{abstract}

	A small and light polystyrene ball is released without initial 
speed from a certain height above the floor. Then, it falls on air. The 
main responsible for the friction force against the movement is the wake 
of successive air vortices which form behind (above) the falling ball, a 
turbulent phenomenon. After the wake appears, the friction force 
compensates the Earth gravitational attraction and the ball speed 
stabilises in a certain limiting value $V_\ell$. Before the formation of 
the turbulent wake, however, the friction force is not strong enough, 
allowing the initially growing speed to surpass the future final value 
$V_\ell$. Only after the wake finally becomes long enough, the ball 
speed {\it decreases} and reaches the proper $V_\ell$.

\end{abstract}

\newpage
 \section{Introduction}

	While teaching Physics at their universities for undergraduate 
students, the authors designed a simple didactic experiment in order to 
exhibit the influence of air flow around falling objects. An unexpected 
behaviour was found, its explanation being beyond undergraduate Physics. 
Previously, a simple, home-made version of the experiment presented here 
was performed, using a staircase, a tape measure and a hand chronometer, 
as follows. A polystyrene ball with diameter $D \approx 2.5{\rm cm}$ and 
mass $m \approx 0.2{\rm g}$ is released from a height $X$ above the 
floor, and the falling time $T$ is measured.

\begin{figure}[!hbt]

\begin{center}
 \hskip-17pt\includegraphics[scale=0.5,angle=-90]{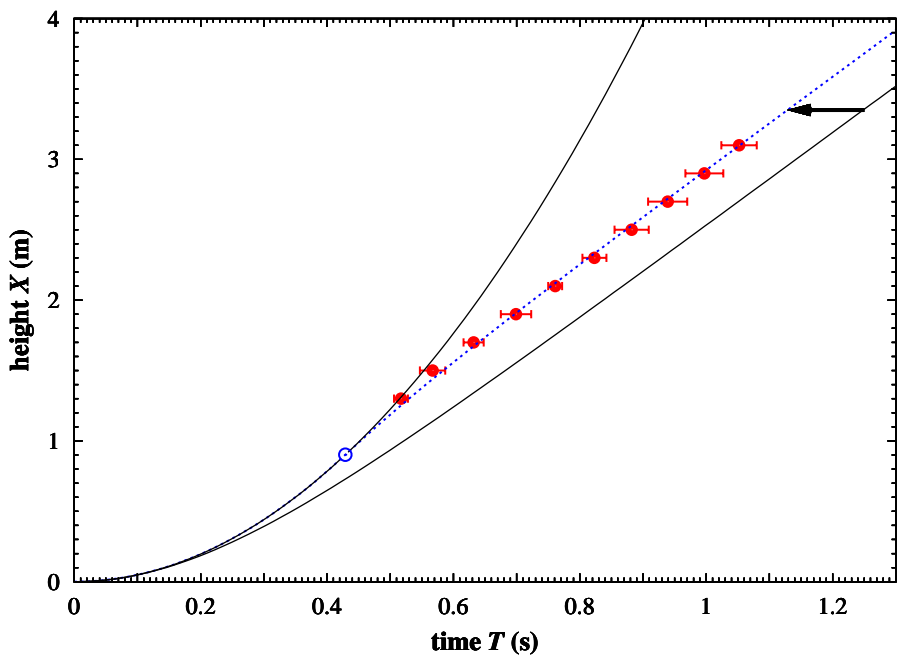}
\end{center}

\caption{Falling time for different heights. The vertical error bars are 
negligible. The horizontal error bar displayed for each experimental 
point (red) is calculated from the statistics of 10 repeated time 
measurements. Using the Earth gravitational field $g = 9.8{\rm m/s^2}$, 
the upper curve is the free fall parabola, $X = g\, T^2/2$, no friction 
at all. The other two curves correspond to an air friction force 
proportional to the squared speed. At the lower curve, this force acts 
during the whole fall, since beginning. At the intermediate, dotted 
curve (blue), the same friction force is turned on {\it only after} an 
adjusted initial transient time (open blue bullet), with no friction 
before that. Note that the slope of this curve has a maximum around 
$\approx 0.4{\rm s}$. Note also that the observed fall is in advance by 
more than $0.1{\rm s}$ (arrow), relative to the lower curve.}

\label{fig1}
\end{figure}

	The procedure is repeated 10 times for each height, in order to 
determine the experimental uncertainty and the consequent error bars. 
The result is displayed by the bullets with horizontal error bars in 
figure 1.

	The traditional free fall model, $X = g\, T^2/2$, obviously does 
not fit the experimental data, perhaps with the exception of the two 
first experimental points obtained for the smallest measured heights. 
Therefore, somehow the influence of the air flow around the ball should 
be taken into account. The two first experimental points positioned near 
the parabola indicate a deviation from the free fall {\it only after} a 
certain transient time, during which the influence of the air around the 
falling ball seems negligible.

	Ruled out the free fall, the next simplest model one can try is 
the also traditional Stokes friction force, $F = 3\pi D \eta V$ for a 
ball running with constant speed $V$ on a viscous fluid with viscosity 
$\eta$. This formula is valid only while the air flow around the ball is 
laminar, without turbulent vortices, which does not correspond to our 
experimental conditions. Even so, we can estimate the order of magnitude 
$F \approx 10^{-5}{\rm N}$ of this force by taking the maximum observed 
speed $V \approx 4{\rm m/s}$, and by using the air viscosity $\eta 
\approx 2 \times 10^{-5}{\rm kg/m.s}$. It is therefore negligible, two 
orders of magnitude smaller than the ball weight $mg \approx 2 \times 
10^{-3}{\rm N}$. So, we can conclude that it is not the Stokes laminar 
flow friction proportional to $V$ which matters within our experiment.

	Indeed, a Reynolds number ${\cal R} \approx 6 \times 10^3$ can 
be estimated for $V \approx 4{\rm m/s}$. (The Reynolds number defined as 
${\cal R} = \rho\, V\, D/\eta$, where $\rho \approx 1.2{\rm kg/m^3}$ is 
the air density, is the main index to evaluate the various possible 
regimes of the air flow around the ball, laminar, turbulent, etc.) For 
$10^3 < {\cal R} < 10^5$, the friction force is experimentally known to 
be proportional to the squared speed within a good accuracy, and also 
much larger than the Stokes laminar prediction. The experiments are 
performed in wind tunnels, with the ball fixed under a constant wind 
speed, which again are not the same conditions of our experiment.

	Besides measuring the friction force, these wind tunnel 
experiments also allow to measure the air velocities in different points 
near the ball at different times. The outcome is the so-called {\it von 
K\'arm\'an vortex street}, a wake of successive air vortices which form 
behind the ball, extending for distances corresponding to hundred 
diameters downstream. The turbulent vortex dynamics is the following. 
One vortex forms near the ball and slowly goes away. When its distance 
from the ball reaches some few diameters, a second vortex forms near the 
ball, which also slowly goes away. Then, a third vortex forms again near 
the ball, and so on. After some time, there is a complete wake of many 
successive vortices behind the ball, the {\it street}.

	By following the movement of one particular vortex, one notes 
that its speed is much smaller than the wind speed itself (far from the 
ball and the wake). Therefore, somehow the ball drags the vortex wake 
behind it. According to the third Newton's law, the reaction force 
exerted by the wake on the ball substantially enhances the friction 
force against the movement, compared with the Stokes laminar case, which 
for that reason is sometimes called {\it drag} force.

	An excellent description of this kind of experiments can be 
found in reference \cite{Feynman}. The knowledge about this problem is 
almost completely obtained from experiments, no first principle theory 
is available. Within the last half century, besides true experiments in 
wind tunnels, numerical experiments were also carried out by solving the 
phenomenological, non-linear Navier-Stokes dynamic equations, a very 
hard numerical task. For a good review in the particular case of a 
sphere, see \cite{Jones}. Almost all human knowledge about the important 
field of fluid dynamics is based exclusively on these two pillars: wind 
tunnel experiments and numerical solution of the phenomenological 
Navier-Stokes equations. Thanks to both, we have flying planes, good 
understanding of the bloodstream, oil pipelines, and other modern 
technologies available to humankind.

	In our particular case of the polystyrene falling ball, the only 
informations we need to keep in mind follow. A) The continuous 
production of successive vortices does not exist at the beginning, since 
the ball is released with initial zero speed. Therefore there is some 
{\it transient} time one must wait before the appearance of the 
turbulent wake and the consequent drag force. B) The resulting {\it 
steady-state} drag force is proportional to the squared speed, within 
the range of Reynolds numbers we are dealing with. The quoted 
steady-state corresponds to the turbulent wake already established. 
Before that, the friction is certainly smaller than its steady-state 
value measured in wind tunnels.

	Besides the experimental data in figure 1, we have also 
estimated the final speed $V_\ell = 3.3{\rm m/s}$, by measuring the time 
for a much larger fall height of $8.9{\rm m}$. This measurement allows 
us to write the (steady-state value for the) friction force as

$$F = \frac{mg}{V_\ell^2}\, V^2\,\,\,\, . \eqno(1)$$

\noindent As commented before, the exponent 2 is valid within the 
interval $10^3 < {\cal R} < 10^5$. The limit ${\cal R} = 10^3$ is soon 
reached by our polystyrene ball, after falling some centimetres, 
allowing equation (1) to be kept during the whole fall. {\it But only as 
an upper limit}, of course, because during the fall beginning, say the 
first meter, the wake-less friction force is somehow smaller than that 
steady-state limit. Equation (1) should be taken as the real drag force 
{\it only after} the vortex wake appears. Indeed, by solving Newton's 
law of motion

$$\frac{{\rm d}V}{{\rm d}T} = g\, [1-(V/V_\ell)^2] \eqno(2)$$

\noindent which includes the friction term $-(V/V_\ell)^2$ during all 
the time, one gets the lower continuous curve in figure 1. In spite of 
equation (1) being already calibrated by the measurement of $V_\ell$, 
this model obviously overestimates the effect of the friction force. The 
overestimation occurs not in the value of the force itself, but because 
the friction term $-(V/V_\ell)^2$ was taken since $t = 0$, too early. 
The consequence is a delay larger than $0.1{\rm s}$ which can be 
observed between the set of experimental points and the lower continuous 
curve in figure 1 (horizontal arrow). Since the delay is too large 
compared with the experimental accuracy, this second model should also 
be ruled out.

	Different from previous works \cite{AJP}, at the beginning of 
the fall, we should replace equation (2) by something else. The simplest 
option is to drop completely the drag term $-(V/V_\ell)^2$ until some 
arbitrarily chosen point, say the open blue bullet in figure 1, 
returning back to equation (2) afterwards. The result is the 
intermediate, dotted blue line in figure 1, which indeed fits well all 
the experimental data (including $V_\ell$).

	The complete absence of friction up to some point is an extreme 
approach, nevertheless realistic as follows. Applying to our ball the 
traditionally known experimental drag coefficient as a function of the 
Reynolds number, we could estimate the (steady-state) friction force $F$ 
and compare it to the weight $mg$. A ratio $F/mg << 10^{-2}$ holds up to 
${\cal R} = 300$ when it reaches $10^{-2}$. For ${\cal R} = 1000$ it is 
$10^{-1}$. For ${\cal R} = 3000$ the steady-state friction $F$ starts to 
approach the weight $mg$. Remembering that this force is only an upper 
limit for the case of our falling ball, it is plausible to consider 
${\cal R} = 3000$ or a little bit above as the point when the turbulent 
wake {\sl lately} appears and develops. Before that, even the upper 
limit of this force is negligible when compared to the ball weight.

	In short, this initially naive experiment shows us a surprising 
effect: the falling ball indeed loses speed! The phenomenon behind this 
effect is much more complex than we imagined within our initial didactic 
purposes. Namely, there is a {\it transient} time during which the long 
{\it von K\'arm\'an vortex street} is still absent. During this 
transient, the friction force acting on the ball is substantially 
smaller than its steady-state value (proportional to $V^2$) measured in 
wind tunnels experiments. Consequently, also during this transient, the 
ball can acquire a speed larger than its final value $V_\ell$ imposed 
only after the wake is already developed.

	In principle, the successive ball speeds could be obtained from 
the same experimental data, by dividing the difference of $\Delta X = 
20{\rm cm}$ between successive heights by the respective differences 
$\Delta T$ between the measured falling times. Indeed, this procedure 
indicates a maximum speed a little bit above $4{\rm m/s}$ near $t = 
0.5{\rm s}$, whereas the final speed $V_\ell = 3.3{\rm m/s}$ is sensibly 
smaller than that. Unfortunately, by calculating differences, the 
accuracy in $\Delta T$ becomes very poor. Therefore, we decided to 
perform a more sophisticated version of this simple experiment, allowing 
us to measure directly the speeds, as described hereafter.

 \section{The Experiment}

	We used a digital camera storing successive snapshots every 
$(1/2000){\rm s}$. Among a total of 4096 snapshots for each fall, we 
sort only 3 of them: snapshot S0, the very first which determines the 
time $t = 0$ when the ball was released; and snapshots S1 and S2, taken 
when the ball is already close to reach the floor. The camera is {\it 
fixed} also near the floor, horizontally directed towards the very end 
of the trajectory, see figure 2.

\begin{figure}[!hbt]

\vskip-15pt
\begin{center}
 \hskip-17pt\includegraphics[scale=0.4]{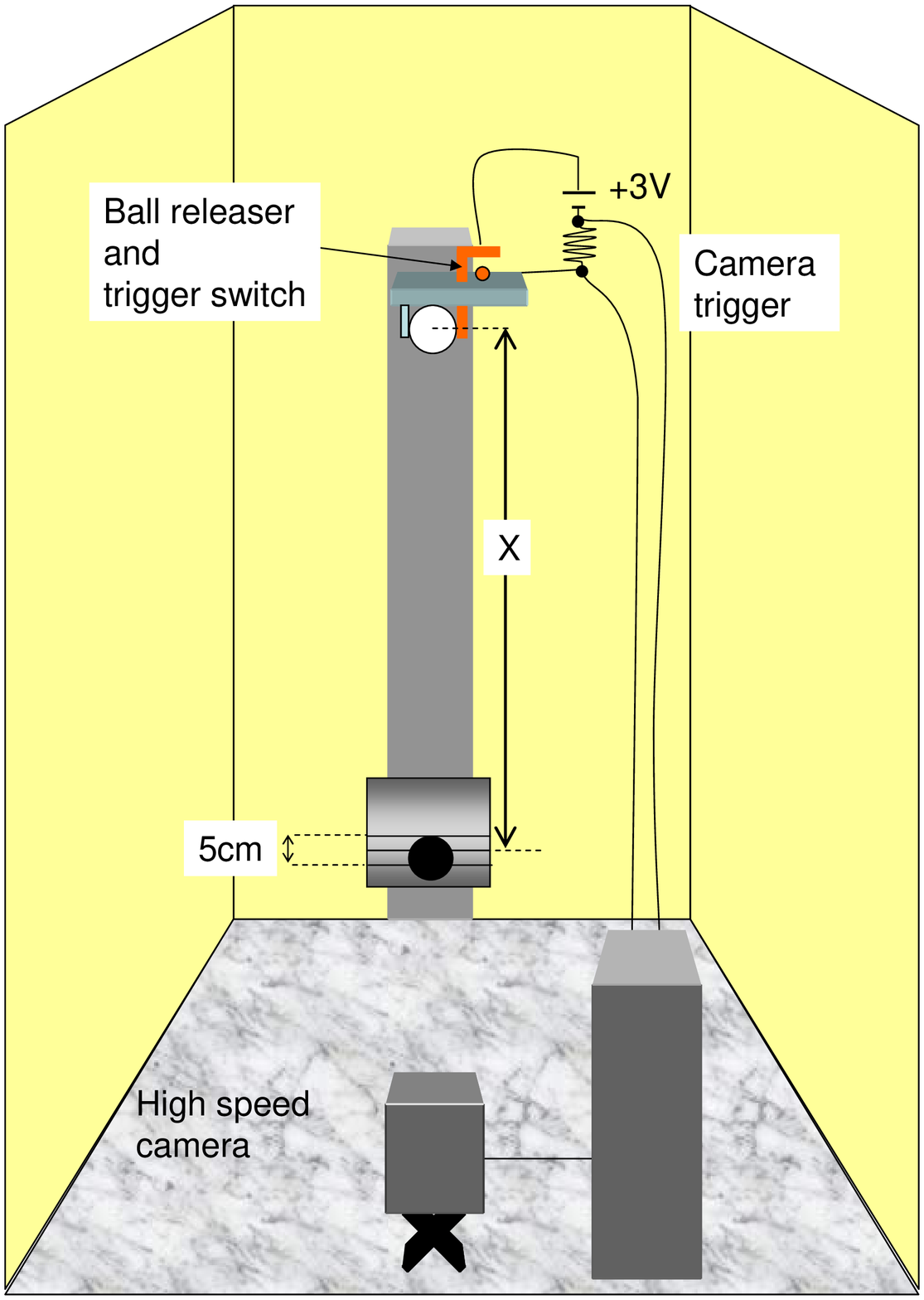}
\end{center}

\caption{Experimental setup.}

\label{fig2}
\end{figure}

	Snapshot S0 does not show the ball just released far above, 
outside the camera visual field. It is used only to determine the 
starting time. The whole system is triggered by a mechanic hook which 
simultaneously releases the ball and sends an electrical signal to the 
camera. Following the recorded images, one after the other on a computer 
screen, one sees nothing during many initial snapshots, while the ball 
is still above the camera visual field. However, the time is being 
counted every $(1/2000){\rm s}$. Suddenly, the ball appears inside the 
visual field, and one can observe the final part of its fall during half 
a hundred snapshots. Among them, snapshots S1 and S2 are chosen as 
described below.

	Three horizontal lines were previously draw on a fixed dull 
glass plate illuminated from behind, in front of which the falling ball 
passes. Some four meters distant, the camera records the successive 
images. The height $X$ between the central line and the point where the 
ball was released is also previously determined. We used two recipes to 
choose snapshots S1 and S2 adopted as measures.

\begin{figure}[!hbt]

\begin{center}
 \hskip-17pt\includegraphics[scale=0.5,angle=-90]{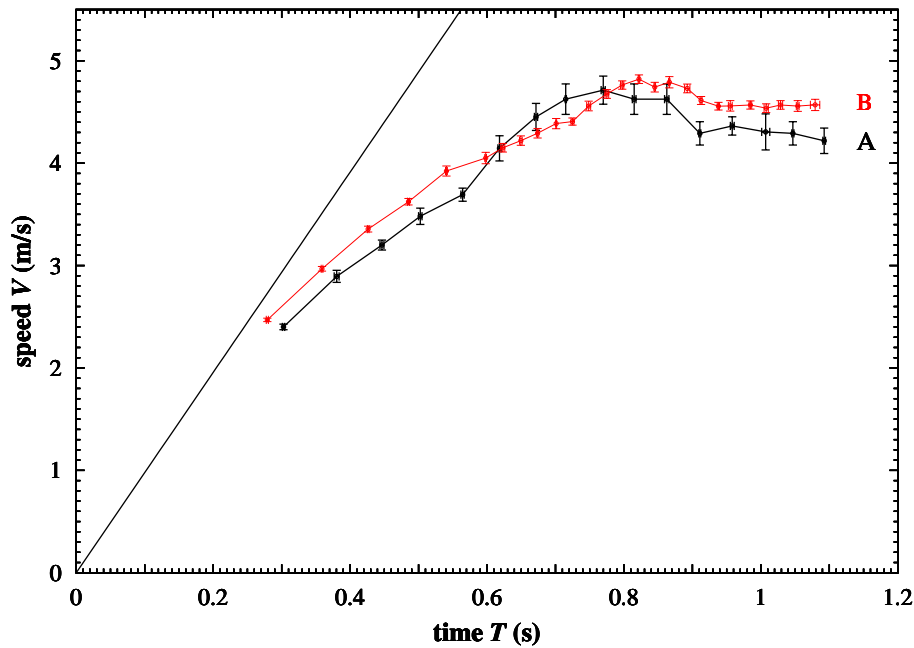}
\end{center}

\caption{Directly measured speeds for a polystyrene ball with mass $m = 
0.22{\rm g}$ and diameter $D = 2.47{\rm cm}$. Two superimposed 
experimental data sets were taken in different days. One with larger 
error bars and less points (A), converging to $V_\ell \approx 4{\rm 
m/s}$, taken during a hot Brazilian day (S\~ao Paulo), temperature 
around $30{\rm ^oC}$. The other, with smaller error bars and more points 
(B), converging to a final speed close to $V_\ell \approx 4.5{\rm m/s}$, 
taken in a Brazilian normal day (S\~ao Paulo again), temperature around 
$25{\rm ^oC}$. The previous measurements of figure 1 were made (with 
another ball of the same size) during a very hot Brazilian day (Rio de 
Janeiro), temperature $\approx 35{\rm ^oC}$, $V_\ell = 3.3{\rm m/s}$ 
measured in a much longer fall of $8.9{\rm m}$. The straight line 
corresponds to the free fall.}

\label{fig3}
\end{figure}

	First, we took snapshot S1 when the ball {\it enters} into the 
central horizontal line, and S2 when the ball {\it exits} the same line. 
In this case, the distance to be divided by the time interval between 
snapshots S1 and S2 is the ball diameter itself. The result is one of 
the 5 speed measurements we repeat for each height, in order to average 
and to determine the error bars, figure 3, curve A.

	Second, we took S1 as the snapshot when the ball {\it exits} the 
upper line, and S2 when it {\it exits} the bottom line. In this case, 
the distance to be divided by the time interval is the fixed distance 
between both lines, in our case $5.00{\rm cm}$. (Also in this case, the 
height $X$ in figure 2 is measured from the upper edge of the not-yet 
released ball.) For small balls ($D < 5{\rm cm}$), this two-line option 
is more accurate than the previously described single-line approach. 
Also, for heights above $1.35{\rm m}$ when the turbulence effects 
appear, we made 10 instead of only 5 repeated measures. Furthermore, 
consecutive heights were taken closer, $10{\rm cm}$ instead of $20{\rm 
cm}$. The result is displayed in figure 3, curve B.

	These experimental results agree with our preliminary conclusion 
concerning the ball {\it decreasing} speed after surpassing the final 
value $V_\ell$.	We have done many other experiments with different 
balls, under the same or different weather conditions. During the same 
day when experiment (B) in figure 3 was performed, we have also observed 
the speed-decreasing behaviour for a smaller ball of diameter $D = 
2.02{\rm cm}$ and mass $m = 0.14{\rm g}$, a little bit denser than the 
ball shown in figure 3. However, for a larger ball of diameter $D = 
3.04{\rm cm}$ and mass $m = 0.35{\rm g}$ the decreasing speed effect was 
hardly visible. Larger yet balls of diameters $5{\rm cm}$ and $7{\rm 
cm}$ definitely do not show the same behaviour.

	The results of another experiment with tiny balls falling in 
water \cite{Mordant} reinforce our conclusion. Although no speed above 
$V_\ell$ was observed for the glass or metallic balls used in this 
experiment, an oscillating speed was observed during the transient. 
First, the speed increases up to a certain value close to the future 
$V_\ell$, then decreases a litle bit, increases again and so on, 
eventually stabilising at $V_\ell$. The authors ``expect the motion of a 
lighter bead to be more influenced by the eventual unsteadiness of its 
wake''\cite{Mordant}. Their interpretation for the speed oscillations is 
also linked to the ``temporal evolution of the particle wake''. They 
also quote that ``the oscillations disappear if the motion is averaged 
over several falls'', showing that ``the events in the wake that are 
responsible for them are not coherent, in the sense that they do not 
occur at fixed times''. We can add that the wake formation is triggered 
by some minor early fluctuations occurred in the air around the ball, 
gradually amplified by the flow non-linear character. The triggering 
fluctuations naturally vary for different fall realisations.

	After reaching $V_\ell$, the ball weight is equal to the drag 
force, leading to the relation $V_\ell^2 = (4\, \rho_{ball}\, g\, 
D)/(3\, C_D\, \rho)$, where $C_D \approx 0.4$ or $0.5$ is the 
experimentally known drag coefficient almost constant for the range of 
Reynolds numbers relevant in our case. This is in perfect agreement with 
our final speeds $V_\ell \approx 4{\rm m/s}$. Moreover, by taking balls 
made of the same material (same $\rho_{ball}$), this relation shows that 
the final speed $V_\ell$ is smaller for smaller balls, also in agreement 
with our experiments. On the other hand, during the fall beginning the 
speed increases according to an acceleration close to $g$, independent 
of the ball size. Therefore, the initially increasing speed of smaller 
balls can surpass their small values of $V_\ell$. But larger balls do 
not, again in agreement with our experiments.

	Another requirement is the ball weight (or density 
$\rho_{ball}$) which should be also small enough in order to allow a 
reasonable difference between the maximum speed and $V_\ell$. For 
another polystyrene ball with the same diameter as those in figures 1 or 
3, but with a smaller mass of $0.14{\rm g}$ (the ball was cut into two 
hemispheres, the inner part was removed and the halves were glued 
again), we have observed the same decreasing speed behaviour, with both 
the maximum speed and $V_\ell$ smaller than those in figure 3. On the 
other hand, two other balls made of a denser polystyrene material ($D = 
2.02{\rm cm}$, $m = 0.29{\rm g}$ and $D = 1.50{\rm cm}$, $m = 0.10{\rm 
g}$) soon reach their maximum speed ($\approx 6{\rm m/s}$ at $\approx 
0.8{\rm s}$ and $\approx 5{\rm m/s}$ at $\approx 0.9{\rm s}$ 
respectively), but the speed decreasing after that is hardly noticeable 
due to the limiting height of $3.35{\rm m}$ inside the laboratory room. 
As $6{\rm m/s}$ and $5{\rm m/s}$ are already close to the expected 
values for their final speeds, probably the decreasing speed effect 
would not appear for these two balls. The direct measurement of $V_\ell$ 
through a much higher height, as we have done for the ball in figure 1, 
is impossible inside the laboratory room. Outside, on the other hand, it 
would be useless because the weather conditions would be modified. (All 
data in figure 1 as well as the value $V_\ell = 3.3{\rm m/s}$ were 
obtained with a hand chronometer, instead of the sophisticated camera, 
at the same day and outside the laboratory room.)

	A third important element is of course the weather itself. 
Different temperatures, atmospheric pressures, air humidity and density 
modify the air viscosity effects. Experiments inside the laboratory room 
with the air conditioner turned on (temperature $\approx 20{\rm ^oC}$, 
air density, humidity and pressure not measured) also did not show the 
decreasing speed effect.

 \section{Phenomenological Support}

	Our experimental result and its interpretation can be summarised 
as follows. Released without initial speed, a small and light enough 
ball (diameter $D \approx 2.5{\rm cm}$ and mass $m \approx 0.2{\rm g}$, 
i.e. made of very low density material with at most $\approx 20$ times 
the air density) presents a maximum speed during its fall on air. Then, 
the speed decreases and reaches a smaller and constant value $V_\ell$, 
but only after a turbulent vortex wake is completely developed behind 
(above) the ball, the so-called {\it von K\'arm\'an vortex street}.

	The crucial feature of our interpretation is the initial absence 
of the turbulent wake behind the ball, while its speed gradually 
increases. Can this scenario be confirmed through the solution of 
Navier-Stokes equations?

	Some difficulties arise. The first one is a non-constant 
Reynolds number ${\cal R} = \rho\, V\, D/\eta$ in our case, different 
from traditional wind tunnel experiments where only the steady-state 
situation is measured with fixed $V$. Suppose we adopt a numerical 
approach, with $\Delta t = 0.0025$ as the discrete time interval, in 
units of $D/V_\ell$: we would need some $6 \times 10^4$ time steps in 
order to follow the first second of the fall. Considering one hour the 
processing computer time for each time step (a plausible estimate with 
single processors), the whole thing would spend 7 complete years.

	This difficulty leads us to adopt some simplifications. First, 
to consider a fixed Reynolds number, say ${\cal R} = 30$ or $1000$, but 
starting the solution at $t = 0$ with the Stokes configuration of 
laminar air flow around the ball, previously obtained with ${\cal R} \to 
0$ (or $V \to 0$). This would correspond not exactly to our experiment, 
but to an imaginary experiment realised in a wind tunnel initially kept 
off (air around the ball at rest) and suddenly turned on with a fixed 
wind speed. Our purpose is to observe only the first snapshots of the 
wake formation, far before the steady-state situation. In particular, we 
are interested in verifying the existence or not of some transient 
regime occurring before the turbulent wake with the endless formation 
of successive vortices appears.

	Another simplification is to replace the ball by a long cylinder 
perpendicular to the air flow, supposing perpendicular to the cylinder 
all velocities in different points of the air around. This trick reduces 
our numerical effort from 3 to 2 dimensions. Indeed, according to many 
real experiments with a cylinder, the vortices are also cylindrical and 
parallel to it, with good accuracy, within the range of Reynolds numbers 
we are dealing with. Also the known experimental plots for the 
steady-state drag force as a function of the Reynolds number are 
completely equivalent, even quantitatively, for a ball or for a 
cylinder. All orders of magnitude are the same. In reality, reference 
\cite{Feynman} shows the experimental plots for a cylinder (see also the 
classical papers \cite{Provansal,Williamson}) and not for a ball, which 
can be found elsewhere (for instance \cite{Nakamura,Jones}).

	Within the unusual but very interesting Feynman formulation, the 
Navier-Stokes equation reads

$$ \frac{\partial{\vec \omega}}{\partial t} = \frac{1}{\cal R}
\nabla^2\vec{\omega} - \vec{\nabla}\times(\vec{\omega}\times\vec{v}) $$

\noindent where $\vec v$ are the air velocities in different positions. 

	The vorticities

$$\vec{\omega} = \vec{\nabla}\times\vec{v}$$

\noindent are auxiliary uni-dimensional vectors parallel to the cylinder. 
They are strategically located at the centres of each square plaquette 
of the $\vec v$ grid used for the numerical solution.

	Moreover, air density fluctuations do not appear in our 
experiment because all speeds are much smaller than the sound speed 
($\approx 330{\rm m/s}$). Therefore we can set

$$\vec{\nabla}\, \bullet\, \vec{v} = 0\,\,\,\, .$$

	The above Navier-Stokes equation is already written in an 
adimensional form, i.e. the cylinder diameter is $D = 1$ and the wind 
speed (far from the cylinder and the wake) is $V = 1$ pointing along the 
$X$ axis. In order to translate the results for normal units, one should 
simply adopt the real $D$ as the length unit and the real $V$ as the 
speed unit. Consequently, the time unit is $D/V$.

	We solved this equation for the following boundary conditions. A 
rectangle is considered between $x = -3$ and $x = +7$, and between $y = 
-2.5$ and $y = 2.5$, divided into $400\times200$ pixels. Outside this 
rectangle, the wind speed is $V = 1$ pointing along $X$. The cylinder 
axis coincides with $x = y = 0$, and the speed is zeroed inside it, $x^2 
+ y^2 \le (1/2)^2$. Between these two boundaries, two-dimensional 
velocities on the 80,000 grid points (the centre of each pixel) were 
determined numerically, through the above equations, as follows. Finite 
difference versions of the differential operators were used with second 
order accuracy, both in time and space. From one time step to the next, 
all $\vec \omega$ were relaxed in order to obey the $\nabla^2$ operator 
in the first equation, under fixed $\vec v$. After each (whole grid) 
$\vec \omega$ relaxation, corrected vectors $\vec v$ were determined 
from the other two equations applied to the fixed $\vec \omega$, by also 
relaxing all $\vec v$ some hundred times. Then, all $\vec \omega$ were 
relaxed again, and so on, until numerical convergence over the whole 
grid.

	As far as we know, this numerical approach is new. We did not 
pay much attention to improve its computer time performance, a task 
postponed to the future. Many other methods are already adopted and 
optimised (mainly by engineers), including some very efficient 
commercial softwares \cite{Jones}. Most of them are applied in order to 
study the steady-state regime. A few recent works treat the transient 
regimes \cite{transient}, although all being very far from the specific 
case of our falling ball.

\begin{figure}[!hbt]

\begin{center}
 \hskip-17pt\includegraphics[scale=0.49,angle=-90]{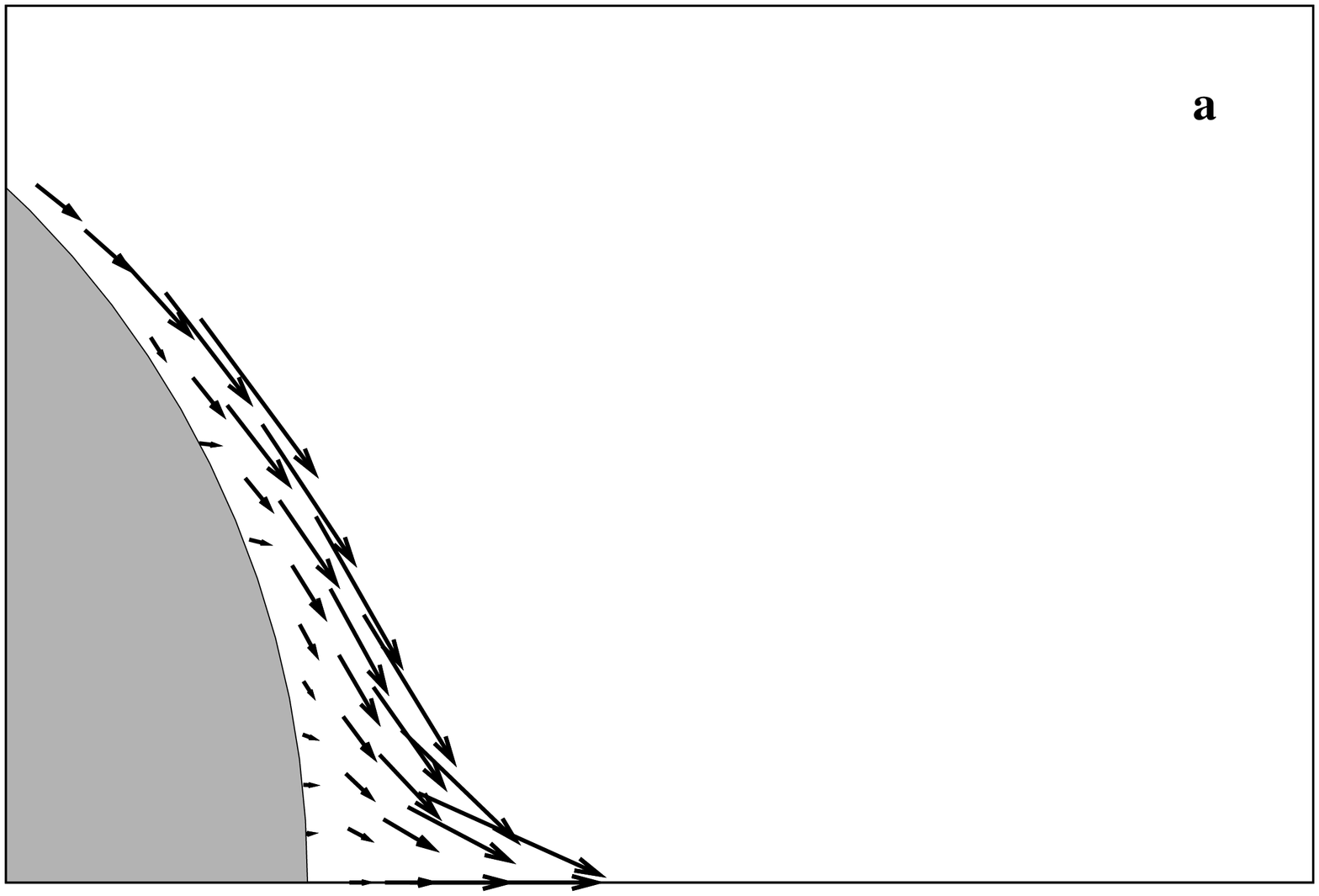}
\end{center}

\label{fig4a}
\end{figure}

\begin{figure}[!hbt]

\begin{center}
 \hskip-17pt\includegraphics[scale=0.49,angle=-90]{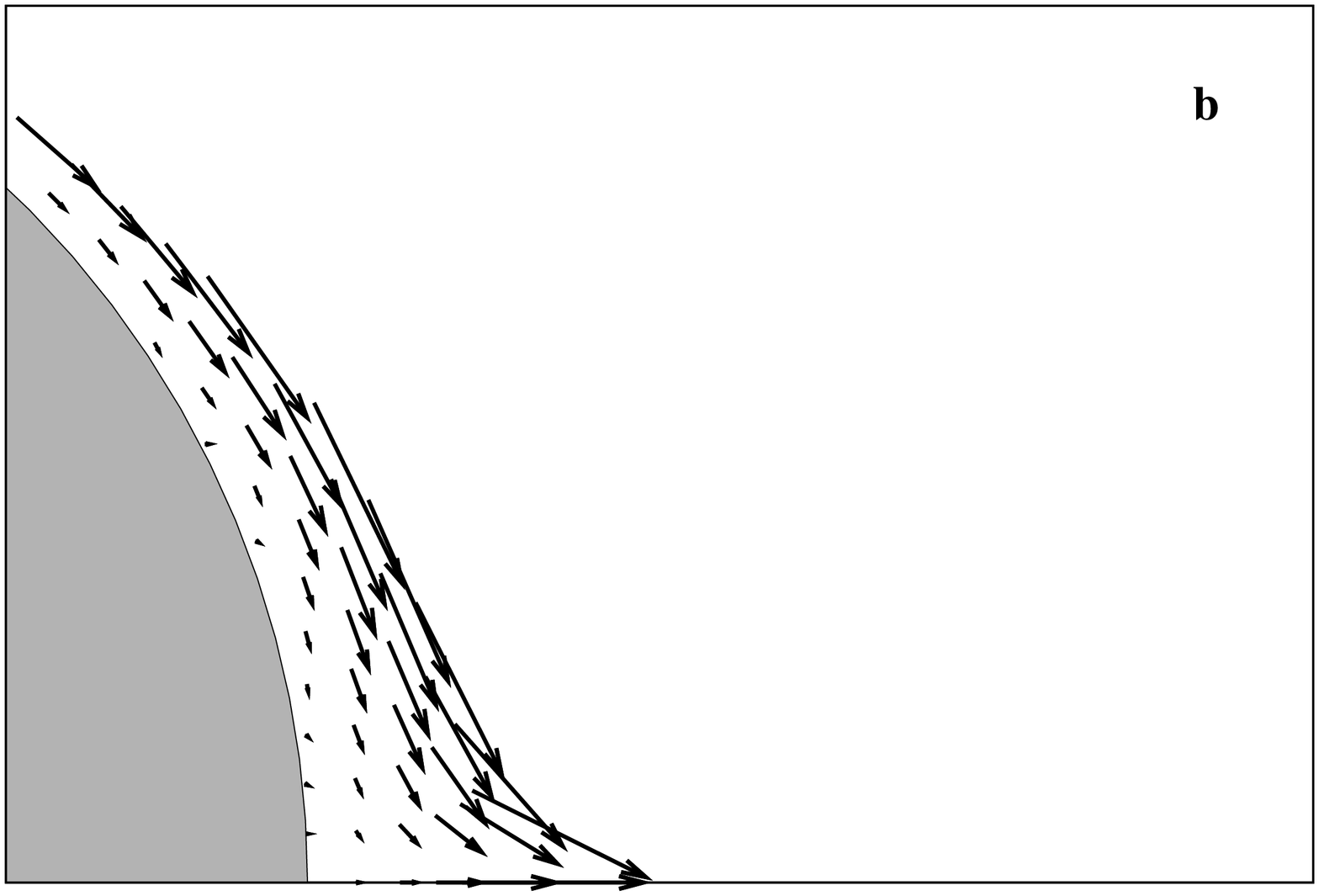}
\end{center}

\label{fig4b}
\end{figure}

\newpage
\begin{figure}[!hbt]

\begin{center}
 \hskip-17pt\includegraphics[scale=0.49,angle=-90]{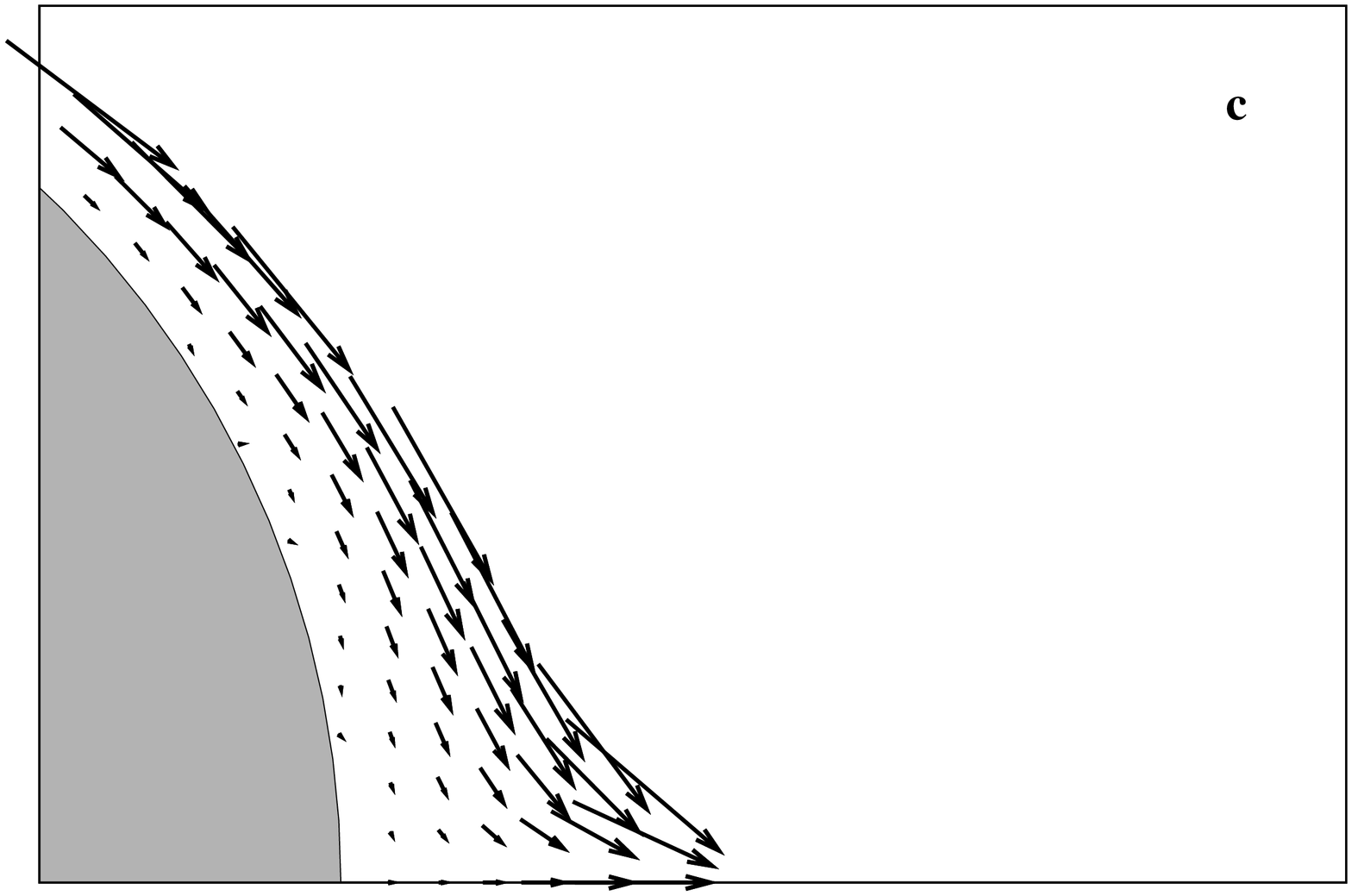}
\end{center}

\label{fig4c}
\end{figure}

\begin{figure}[!hbt]

\begin{center}
 \hskip-17pt\includegraphics[scale=0.49,angle=-90]{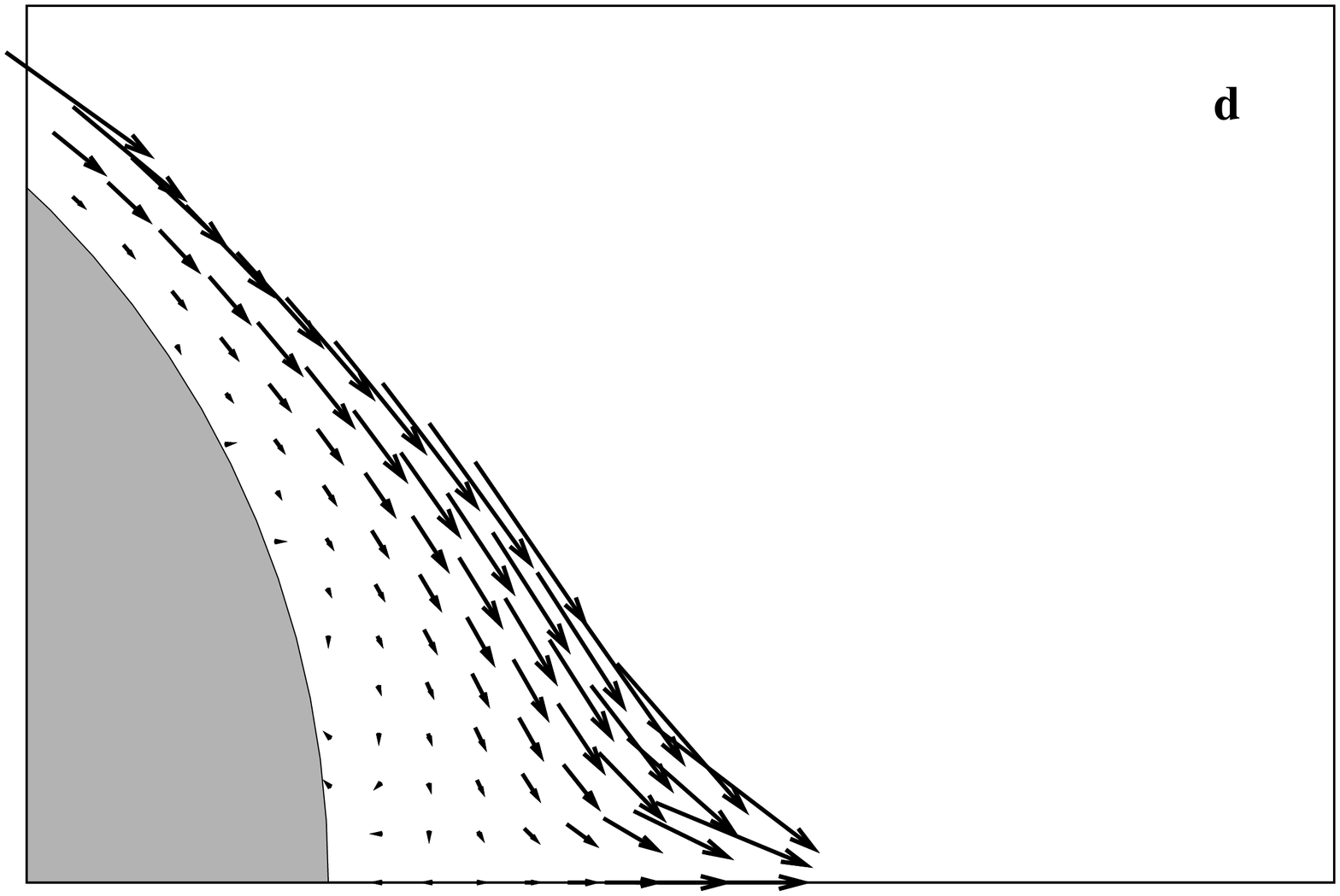}
\end{center}

\label{fig4d}
\end{figure}

\newpage
\begin{figure}[!hbt]

\begin{center}
 \hskip-17pt\includegraphics[scale=0.49,angle=-90]{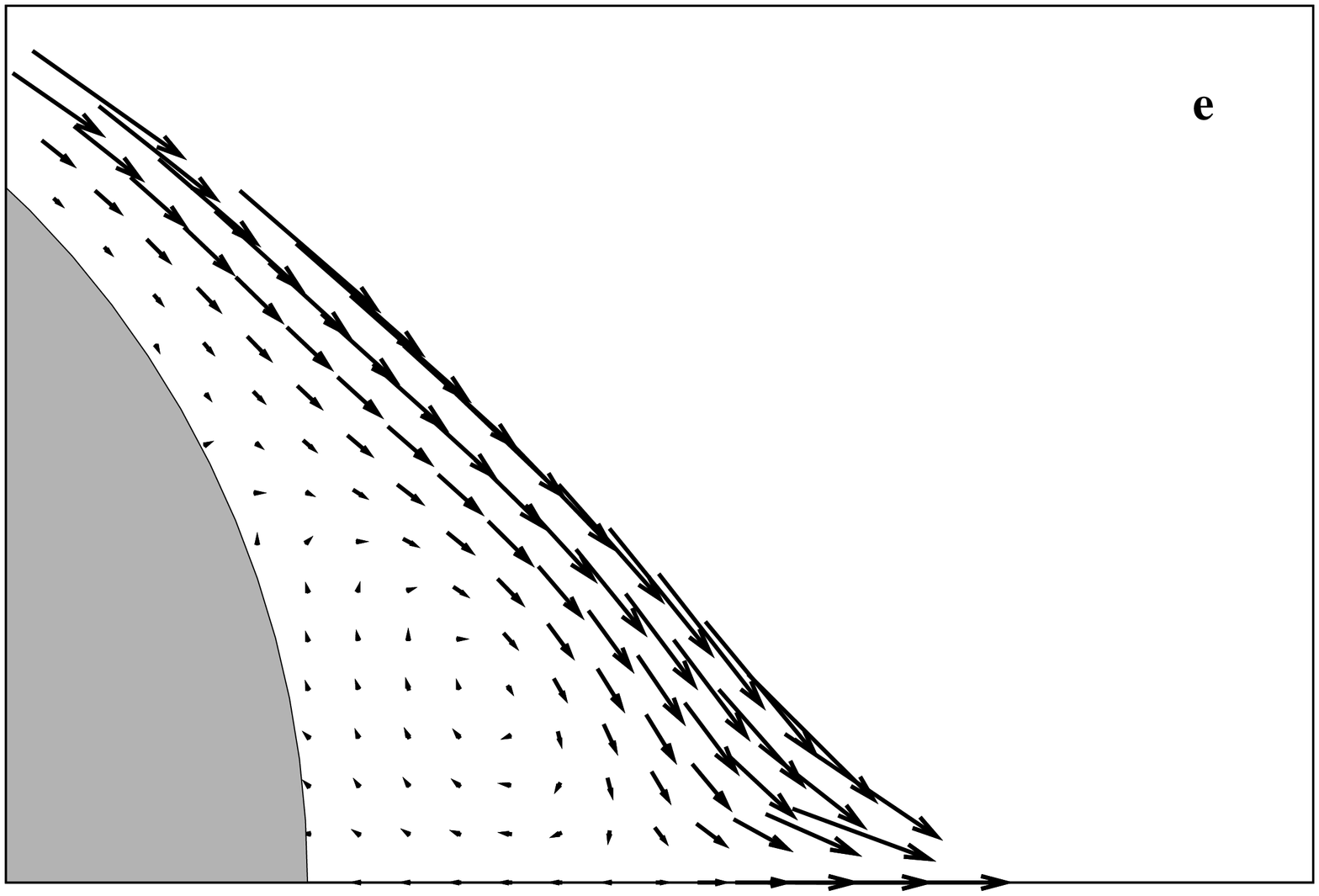}
\end{center}

\label{fig4e}
\end{figure}

\begin{figure}[!hbt]

\begin{center}
 \hskip-17pt\includegraphics[scale=0.49,angle=-90]{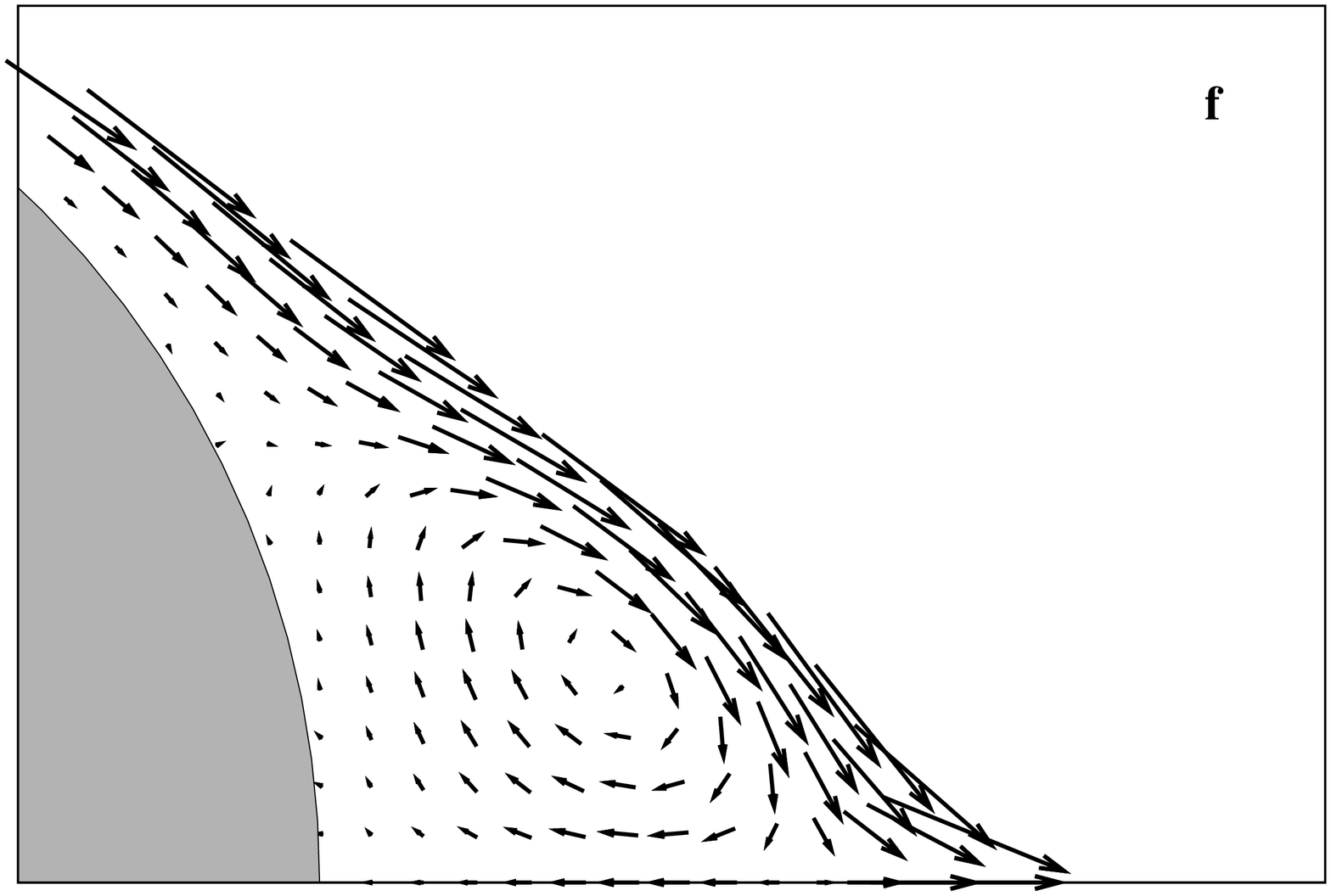}
\end{center}

\label{fig4f}
\end{figure}

\newpage
\begin{figure}[!hbt]

\begin{center}
 \hskip-17pt\includegraphics[scale=0.49,angle=-90]{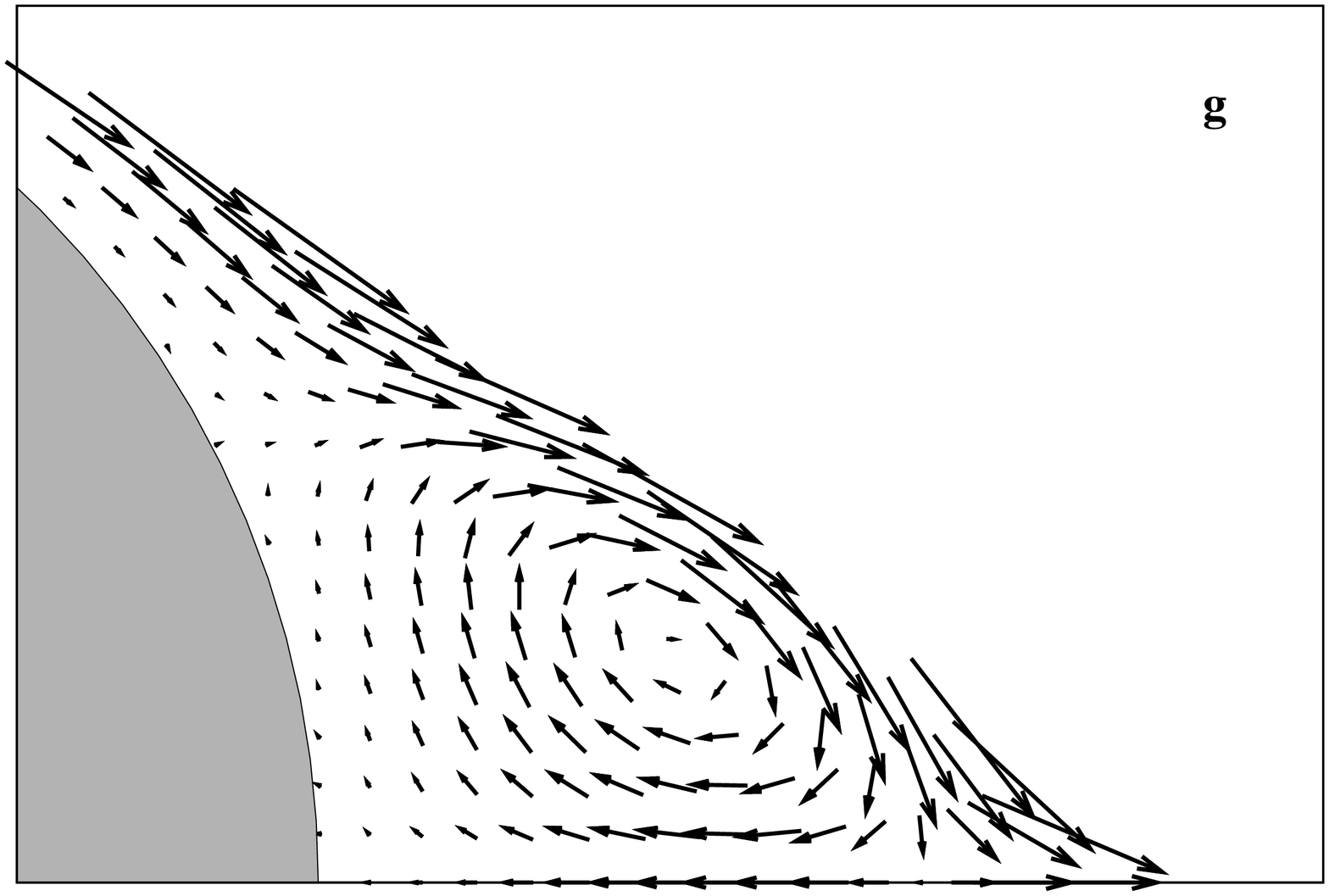}
\end{center}

\label{fig4g}
\end{figure}

\begin{figure}[!hbt]

\begin{center}
 \hskip-17pt\includegraphics[scale=0.49,angle=-90]{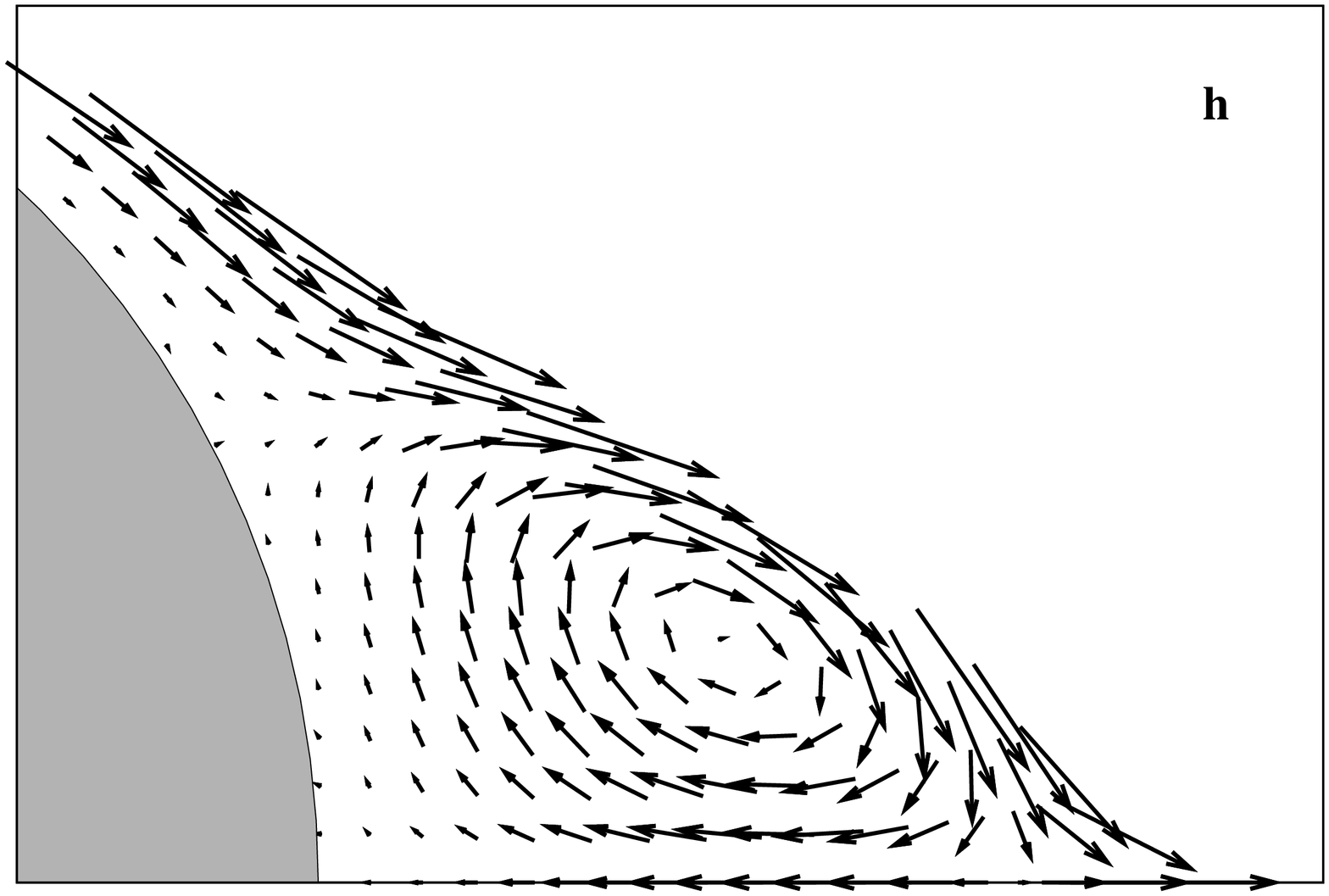}
\end{center}

\label{fig4h}
\end{figure}

\newpage
\begin{figure}[!hbt]

\begin{center}
 \hskip-17pt\includegraphics[scale=0.49,angle=-90]{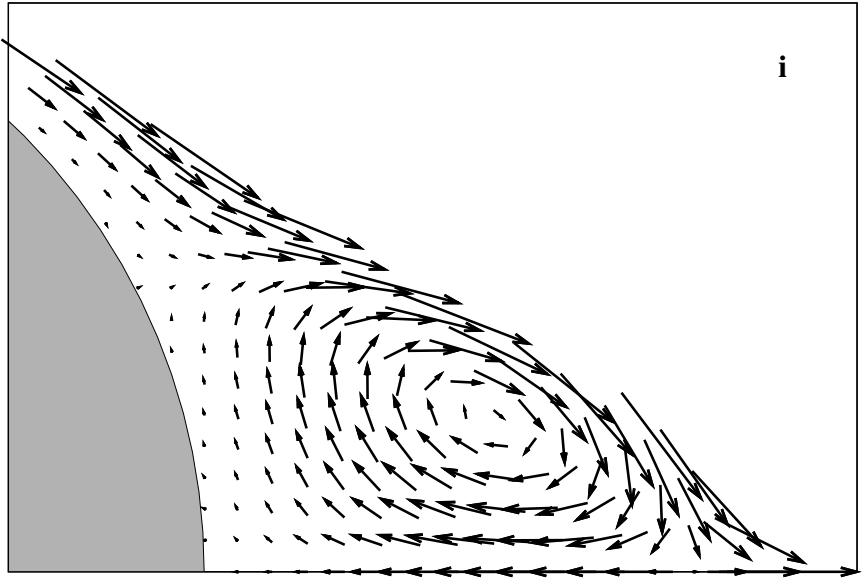}
\end{center}

\caption{Successive snapshots showing the vortex formation behind a 
fixed cylinder (part of its grey back edge is shown on bottom left). 
Arrows represent local air velocities. Only a very small region behind 
the cylinder is shown, half a diameter downstream along the $X$ axis. 
Also, only points above the $X$ axis are shown, below it all velocities 
are mirror symmetric. Only very small speeds $V < 0.007$ are 
represented. Downstream, much larger speeds $V \approx 1$ are omitted 
for clarity in the white (top right) part of the figures. The initial 
configuration (a) corresponds to the Stokes laminar flow ${\cal R} \to 
0$ previously determined. Then, at $t = 0$, the Reynolds number is 
suddenly turned on to ${\cal R} = 30$. Time runs from figures (a) to 
(b), (c), etc, showing the gradual formation of the vortex we will call 
UP. The $X$ axis acts as a mirror, a second, symmetric vortex DOWN (not 
shown) spinning in the other sense is also simultaneously formed below 
each figure. For this small Reynolds number, these two vortices become 
eventually stuck at the position shown in the last figure, with 
negligible fluctuations around. For larger Reynolds numbers and after 
some time, however, each of these same two initial vortices alternately 
bifurcate, forming the long {\it von K\'arm\'an vortex street} (see 
text).}

\label{fig4i}
\end{figure}

	For a nice animation of the flow past a sphere, see 
\cite{animation}: the initial single vortex is a torus symmetric around 
the movement axis, which forms behind but near the ball. Afterwards, 
this symmetry is broken and the long turbulent wake with successive 
vortices develops, no longer torus-like. Returning back to the 
cylindrical symmetry we adopted in our numerical solution, figures 4a, 
4b $\dots$ 4i show consecutive snapshots of the result for ${\cal R} = 
30$.

	We have done the same for ${\cal R} = 1000$ (not shown). The 
beginning is the same, two symmetric vortices, UP and DOWN, appear. The 
most noticeable difference is the (horizontally) stretched form of these 
vortices, instead of the rounded form shown in figure 4. Also, due to 
the stretching, they don't remain stuck forever. After some time, one of 
them (UP in our computer run) bifurcates in two, forming a no longer 
symmetric wake with 3 vortices. Later, vortex DOWN also bifurcates, 
leading to 4 vortices, and so on. After some time, a very long wake of 
alternate vortices is dragged by the cylinder. Then, its contribution to 
the drag force becomes important.

	In short, there are two different steady-state regimes. For 
small enough Reynolds numbers, only two symmetric vortices form and 
eventually stabilise at a fixed distance behind the cylinder. For large 
Reynolds numbers, instead of two vortices at rest, a many-vortices long 
wake is formed downstream, continuously fed with new vortices 
periodically appearing near the cylinder.

	As a technical detail, for ${\cal R} = 1000$ we have modified a 
little bit the boundary conditions at $x = 7$. Instead of fixing $V = 1$ 
outside, we copy the rightmost column of velocities still inside the 
rectangle to the next column already outside. Then, for the next time 
step, we fix this column with velocities which are weighted averages 
between this copy and $V = 1$, taking the averaging weights for the copy 
proportional to the distance to the $X$ axis (zero weight at $y = \pm 
2.5$, unit weight at $y = 0$). This artificial trick allows the wake to 
escape through the back door, but certainly also introduces some unknown 
perturbation on the wake after many vortices were already escaped. Even 
so, we could observe the formation of many new vortices, not yet in the 
steady state regime within our limited computer time. This unknown 
influence of the boundary conditions, however, is not important here 
because we are interested only in the beginning of the process, with 
only the first two symmetric vortices, when one of them bifurcates still 
inside the rectangle ($x \approx 3$ for ${\cal R} = 1000$).

	For a would-be gradually increasing-${\cal R}$ numerical 
solution, we can foresee the following scenario. First, only two 
symmetric vortices stabilises for a while into a position initially very 
near the cylinder. Then, as the Reynolds number increases, these 
vortices become stretched, and their centres reach regions farther and 
farther behind the cylinder, but still symmetric and at rest (completely 
dragged by the cylinder). Suddenly, a {\it rupture} occurs: one of the 
vortices bifurcates in two, then the other, and so on, triggering the 
construction of a continuously-fed turbulent wake of vortices slowly 
moving downstream. Only after this rupture, the friction force starts to 
become important.

	This scenario is {\it qualitatively} compatible with our 
polystyrene falling ball experiment. Although the many-vortices wake is 
already observed in our numerical solution for a {\it fixed} ${\cal R} = 
1000$, we could expect the triggering rupture to occur somewhat later 
within the (unfeasible) gradually increasing-${\cal R}$ solution, where 
the first two symmetric vortices would have time enough to (meta) 
stabilise themselves under the previously smaller Reynolds numbers. 
Slow, adiabatic stretching is naturally more resistant against rupture 
than the sudden stretching we tested by switching ${\cal R} = 0$ 
directly to ${\cal R} = 1000$ at $t = 0$. So, the scenario obtained from 
the phenomenological Navier-Stokes equation can be also {\it 
quantitatively} compatible with our experiment.

 \section{Conclusions}

	By measuring the successive speeds of a polystyrene ball falling 
on air, initially released from a certain height, we could observe a 
completely unexpected behaviour: the ball reaches a {\it maximum} speed, 
then {\it breaks}, and finally a limit speed $V_\ell$ is stabilised, as 
expected. We found two conditions as necessary in order to observe this. 
First, the ball should be light enough in order to allow a small value 
for $V_\ell$ (in our case, $V_\ell \approx 4{\rm m/s}$). Second, the 
ball should be small enough (in our case, diameter $D \approx 2.5{\rm 
cm}$ or less).

	Our interpretation is based on the transient time before the 
development of the so-called {\it von K\'arm\'an vortex street} of air 
vortices behind (above) the ball. This long wake of vortices (some 
hundred diameters) is the responsible for the drag force which 
eventually compensates the gravitational attraction exerted by Earth on 
the ball, providing its final constant speed $V_\ell$. Before the wake 
development, under a smaller dragg force, the ball can reach speeds 
larger than $V_\ell$, because this limit will be imposed only after the 
``infinite'' wake is already developed.

	Numerical solutions of the phenomenological Navier-Stokes 
equations for a cylinder give support to our interpretation. The 
turbulent wake consists of the periodic creation of new vortices which 
slowly move downstream one after the other, the same scenario observed 
in steady-state wind tunnel experiments. However, starting with zero 
speed, our numerical solutions show another transient regime holding 
before the wake formation. First, just two symmetric vortices form and 
become stuck for a while, characterising the quoted transient. Suddenly, 
one of them bifurcates, forming a third vortex, then the other suffers 
the same process, starting the formation of the eventual many-vortices 
wake.

	For spheres instead of cylinders, a single torus-like vortex 
forms behind the sphere at first, and stays dragged by the sphere for a 
while. Suddenly, it breaks down into a sequence of successive, no-longer 
torus-like vortices which go away along the street. We deal with a {\sl 
rupture} phenomenon, a transition between an initial regime where a 
symmetric vortex stays for a while at rest relative to the ball, towards 
a second regime where a sequence of successive non-symmetric vortices 
move away from the ball.

	Besides the speed decreasing curious behaviour, maybe this naive 
experiment could shed light in some {\it transient} and {\it hysteresis} 
phenomena occurring within fluid dynamics, a subject not very well 
known.

\end{document}